# The Impact of Rising Ocean Acidification Levels on Fish Migration

**Asuna P. Gilfoyle** - President & Chief Innovation Officer

**Willow A. Baird** - Contributing Researcher

// Gilfoyle Philanthropies

# Gilfoyle Philanthropies

Gilfoyle Philanthropies is an organization dedicated to advancing our understanding of artificial intelligence, ocean conservation, public health, and socioeconomic inequity through leading research and targeted grants. Through our research we have taken steps towards developing ethical AI programs, furthering our knowledge of the ocean, and promoting critical areas of public health.

**Introduction:**
Ocean acidification, a direct consequence of increased carbon dioxide (CO2) emissions, has emerged as a critical area of concern within the scientific community. The world's oceans absorb approximately one-third of human-caused CO2 emissions, leading to chemical reactions that reduce seawater pH, carbonate ion concentration, and saturation states of biologically important calcium carbonate minerals. This process, known as ocean acidification, has far-reaching implications for marine ecosystems, particularly for marine organisms such as fish, whose migratory patterns are integral to the health and function of these ecosystems.

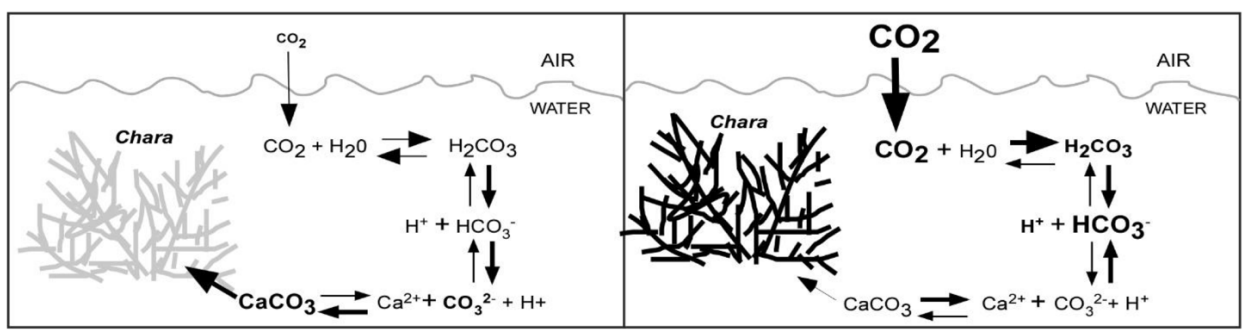

Figure 1: Calcification of *Chara* in presence of low and high amounts of CO2.

Fish migration, a crucial ecological phenomenon, is influenced by a variety of environmental factors. These migratory patterns are not only essential for the survival of individual species but also have significant implications for the biodiversity and resilience of marine ecosystems. Changes in these patterns due to environmental stressors can disrupt the balance of marine ecosystems, affecting predator-prey relationships, competition for resources, and overall biodiversity.

The purpose of this research is to delve into the effects of rising ocean acidification levels on fish migration. While the impacts of ocean acidification on calcifying organisms like shellfish and corals have been extensively studied, its effects on fish, particularly concerning their migratory patterns, is an area that warrants further exploration.

Recent studies have begun to shed light on the physiological impacts of ocean acidification on fish, including effects on sensory function, metabolism, and reproductive health. However, the potential behavioral impacts, including changes to migration patterns, remain poorly understood. This research aims to bridge this knowledge gap and contribute to our understanding of how fish populations may respond to the ongoing acidification of our oceans.

Understanding the impacts of ocean acidification on fish migration is not only crucial from an ecological perspective but also has significant socio-economic implications. Many communities around the world rely heavily on fisheries for their livelihood and food security. Changes in fish migration patterns due to ocean acidification could potentially disrupt these fisheries, leading to economic hardship and threatening food security.

In conclusion, the study of the impacts of rising ocean acidification levels on fish migration is a critical research area that holds significant implications for our understanding of marine ecosystems, fisheries management, and climate change adaptation strategies. Through this research, we aim to contribute valuable insights to this important field of study and inform strategies to mitigate the impacts of ocean acidification.

## 2. Background

2.1 Explanation of the Carbon Cycle

The carbon cycle is a complex series of processes through which all of the carbon atoms in existence rotate. The same carbon atoms in your body today have been used in countless other molecules since time began. The wood burned just a few decades ago could have produced carbon dioxide which through photosynthesis became part of a plant. When you eat that plant, the same carbon from the wood is now in you (BBC, 2014).

The global carbon cycle operates through a variety of response times and mechanisms, from seconds to millions of years, and involves the exchange of carbon between the atmosphere, the oceans, and the terrestrial biosphere. The cycle is usually thought to include the following major reservoirs of carbon interconnected by pathways of exchange: the atmosphere, the terrestrial biosphere (which is usually divided into vegetation and detritus), soils, fossil fuels, the oceans (which include dissolved inorganic carbon and living and non-living marine biota), and sediments (including fossil fuels, freshwater systems, and non-living organic material) (NASA, 2011).

2.2 Role of Oceans in the Carbon Cycle

Oceans play a vital role in the global carbon cycle. They are the largest active carbon sink on Earth, absorbing more than a quarter of the CO2 that humans put into the air.

Oceans absorb atmospheric CO2 through physical and biological processes. In the physical process, CO2 dissolves in the surface water. Afterwards, the ocean's overturning circulation distributes it. The biological process, known as the "biological pump", involves the uptake of CO2 by marine plants in the sunlit surface waters, its conversion to organic carbon via photosynthesis, and the transport of this organic carbon to the deep ocean (NOAA, 2020).

### 2.3 Process of Ocean Acidification

Ocean acidification is a direct result of the increased absorption of CO2 by the oceans. When CO2 is absorbed by seawater, a series of chemical reactions occur resulting in the increased concentration of hydrogen ions. This increase causes the seawater to become more acidic and causes carbonate ions to be relatively less abundant.

Carbonate ions are an important building block of structures such as sea shells and coral skeletons. Decreases in carbonate ions can make building and maintaining these structures difficult for calcifying organisms such as oysters, clams, sea urchins, shallow water corals, deep sea corals, and calcareous plankton. These changes in carbonate chemistry can affect the behavior of non-calcifying organisms as well. Certain fish's ability to detect predators is decreased in more acidic waters. When these organisms are at risk, the entire food web may also be at risk (NOAA, 2020).

In conclusion, the carbon cycle and the role of oceans in this cycle are fundamental to understanding the process of ocean acidification. As CO2 levels in the atmosphere continue to rise due to human activities, the role of the ocean as a carbon sink becomes increasingly important. However, this also leads to increased ocean acidification, posing significant threats to marine ecosystems.

## 3. Impacts of Ocean Acidification on Fish Migration

Ocean acidification, a consequence of increased carbon dioxide (CO2) absorption by the oceans, has significant implications for marine life, including fish. This section explores the impacts of ocean acidification on fish migration, focusing on various aspects such as changes in migratory patterns, physiological effects, and the broader ecological implications.

### 3.1 Changes in Migratory Patterns

Fish migration is a complex process influenced by a variety of environmental factors. Recent research suggests that ocean acidification could alter these migratory patterns. For instance, changes in water chemistry can affect the olfactory cues that many fish species rely on for navigation. Disruptions to these cues could potentially lead to changes in the timing, routes, and success of fish migration (The Journal of Wildlife Management, 2022).

3.2 Physiological Effects
Ocean acidification can also have direct physiological effects on fish, which may in turn impact their migratory behaviors. For example, increased acidity can affect fish's metabolic rates, sensory functions, and reproductive health. These physiological changes could potentially influence the energy availability for migration, alter the sensory cues used for navigation, and affect the timing of reproductive migrations (International Cooperation through Regional Fisheries Management Organizations, 2022).

3.3 Ecological Implications
Changes in fish migration patterns due to ocean acidification can have broader ecological implications. Fish migrations are integral to the functioning of marine ecosystems, influencing predator-prey dynamics, competition for resources, and biodiversity. Alterations to these patterns can disrupt the balance of marine ecosystems, with potential cascading effects on the food web (Anthropogenic Environmental Impacts on Coral Reefs in the Western and South-Western Pacific Ocean, 2022).

3.4 Socio-Economic Implications
Beyond the ecological impacts, changes in fish migration due to ocean acidification can also have significant socio-economic implications. Many coastal communities around the world rely heavily on fisheries for their livelihood and food security. Changes in fish migration patterns could disrupt these fisheries, leading to economic hardship and food security issues (Integration of ocean-based adaptation and mitigation actions into regional and national climate policies in Africa, 2022).

In conclusion, the impacts of ocean acidification on fish migration are multi-faceted, affecting not only the fish themselves but also the broader marine ecosystems and human societies that rely on them. Further research in this area is crucial for understanding these impacts and developing effective strategies for managing and mitigating them.

## 4. Mitigation and Adaptation Strategies

As the impacts of ocean acidification on fish migration become increasingly apparent, it is crucial to develop and implement effective mitigation and adaptation strategies. These strategies must be multi-faceted, addressing not only the direct impacts on fish but also the broader ecological and socio-economic implications.

### 4.1 Reducing CO2 Emissions

The primary driver of ocean acidification is the increase in atmospheric CO2 due to human activities, particularly the burning of fossil fuels. Therefore, a key mitigation strategy is to reduce these emissions. This can be achieved through a variety of means, including transitioning to renewable energy sources, improving energy efficiency, and implementing carbon capture and storage technologies.

### 4.2 Enhancing Ocean Resilience

Another important strategy is to enhance the resilience of marine ecosystems to ocean acidification. This can be achieved through measures such as protecting and restoring key habitats, reducing other stressors like overfishing and pollution, and promoting biodiversity. For example, marine protected areas (MPAs) can provide a refuge for fish and other marine life from fishing pressures, allowing populations to recover and potentially increasing their resilience to ocean acidification.

### 4.3 Monitoring and Research

Ongoing monitoring and research are crucial for understanding the impacts of ocean acidification on fish migration and for informing management strategies. This includes monitoring changes in ocean chemistry and fish migration patterns, researching the physiological responses of fish to ocean acidification, and developing predictive models to forecast future changes.

### 4.4 Socio-Economic Adaptation

Given the potential socio-economic impacts of changes in fish migration due to ocean acidification, it is also important to develop adaptation strategies for affected communities. This could include diversifying livelihoods to reduce reliance on fisheries, developing sustainable aquaculture practices, and providing social safety nets for those affected.

In conclusion, while the challenges posed by ocean acidification are significant, there are a range of potential strategies for mitigating its impacts and adapting to the changes it brings. Implementing these strategies will require concerted effort from all sectors of society, from individual citizens and communities to governments and international organizations.

## 5. Conclusion

Ocean acidification, a consequence of increased carbon dioxide (CO2) emissions, presents a significant challenge for marine ecosystems and the human societies that depend on them. This research has focused on one particular aspect of this challenge: the impacts of ocean acidification on fish migration.

Our exploration of this topic has revealed that these impacts are multi-faceted, affecting not only the migratory patterns of fish but also their physiological health. These changes, in turn, have broader implications for marine ecosystems, disrupting predator-prey dynamics, competition for resources, and overall biodiversity. Beyond the ecological impacts, changes in fish migration patterns can also have significant socio-economic implications, particularly for communities that rely on fisheries for their livelihood and food security.

In response to these challenges, we have outlined a range of potential mitigation and adaptation strategies. These include reducing CO2 emissions, enhancing the resilience of marine ecosystems, conducting ongoing monitoring and research, and developing socio-economic adaptation strategies for affected communities. Each of these strategies has its own set of challenges and opportunities, and their successful implementation will require concerted effort from all sectors of society.

In conclusion, while ocean acidification poses significant challenges, it also presents opportunities for innovation, collaboration, and resilience. By deepening our understanding of the impacts of ocean acidification on fish migration, we can not only help to protect these vital creatures and the ecosystems they inhabit, but also build a more sustainable and resilient future for all. As we move forward, it will be crucial to continue researching this important topic, developing effective strategies for mitigation and adaptation, and working together to implement these strategies on a global scale.

# References

"Factors Affecting Enzyme Action - What Happens during Photosynthesis? - OCR 21st Century - GCSE Combined Science Revision - OCR 21st Century - BBC Bitesize." BBC News. Accessed June 1, 2023. http://www.bbc.co.uk/bitesize/guides/z9pjrwx/revision/3.

"The Carbon Cycle." NASA. Accessed May 10, 2023. https://earthobservatory.nasa.gov/features/CarbonCycle.

Hilderbrand, G.V., White, D., Jr.,, Newman, B. and Krausman, P.R. (2022), The Journal of Wildlife Management is confronting the influences of climate change on wildlife. J Wildl Manag, 86: e22293. https://doi.org/10.1002/jwmg.22293

Molenaar, Erik Jaap. "The Concept of 'Real Interest' and Other Aspects of Co-Operation through Regional Fisheries Management Mechanisms." The International Journal of Marine and Coastal Law 15, no. 4 (2000): 475–531. https://doi.org/10.1163/157180800X00226.

John Morrison, R., and W. G. L. Aalbersberg. "Anthropogenic Environmental Impacts on Coral Reefs in the Western and South-Western Pacific Ocean." In Coral Reefs of the Western Pacific Ocean in a Changing Anthropocene, edited by Jing Zhang, Thamasak Yeemin, R. John Morrison, and Gi Hoon Hong, 7–24. Coral Reefs of the World. Cham: Springer International Publishing, 2022. https://doi.org/10.1007/978-3-030-97189-2_2.

Adewumi, I.J., Ugwu, D.O., & Madurga-Lopez, I. 2022. Integration of ocean-based adaptation and mitigation actions into regional and national climate policies in Africa. Archibald, S.A., Pereira, L.M., Coetzer, K.L., editors.
Future Ecosystems for Africa (FEFA), University of the Witwatersrand: Johannesburg, 157pp.

# Gilfoyle Philanthropies

**gilfoyle.org**

**About Gilfoyle Philanthropies**

Gilfoyle Philanthropies is an organization dedicated to advancing our understanding of artificial intelligence, ocean conservation, public health, and socioeconomic inequity through leading research and targeted grants.

**Board of Directors**

**Asuna P. Gilfoyle**
President & Chief Innovation Officer

**M.K. Matthews**
Director

**Michael "Mike" Mannino**
Director

**Doug D. Wood**
Director